\documentclass[showpacs,preprintnumbers,eqsecnum,nofootinbib,,twocolumn,prd]{revtex4-1}   

\usepackage{graphicx}
\usepackage{latexsym}
\usepackage{amsmath}
\usepackage{amssymb}

\def\beq{\begin{equation}}
\def\eeq{\end{equation}}

\begin{document}

\title{Conservative second-order gravitational self-force on circular orbits \\ and the effective one-body formalism}

\author{Donato \surname{Bini}$^1$}
\author{Thibault \surname{Damour}$^2$}

\affiliation{$^1$Istituto per le Applicazioni del Calcolo ``M. Picone,'' CNR, I-00185 Rome, Italy\\
$^2$Institut des Hautes Etudes Scientifiques, 91440 Bures-sur-Yvette, France}

\date{\today}

\begin{abstract} 
We consider Detweiler's redshift variable $z$ for a nonspinning mass $m_1$ in circular motion (with orbital frequency 
$\Omega$) around a nonspinning mass $m_2$. We show how the combination of effective-one-body (EOB) theory with the first law of binary 
dynamics allows one to derive a simple, exact expression for the functional dependence of $z$ on the 
(gauge-invariant) EOB gravitational potential $u=(m_1+m_2)/R$.
We then use the recently obtained high-post-Newtonian(PN)-order knowledge of the main EOB radial potential $A(u ; \nu)$ [where $\nu= m_1 m_2/(m_1+m_2)^2$] to decompose the
second-self-force-order contribution to the function $z(m_2 \Omega, m_1/m_2)$ into a known part (which goes beyond the 4PN
level in including the 5PN logarithmic term, and the 5.5PN contribution), and an unknown one [depending on the  yet
unknown, 5PN, 6PN, $\ldots$, contributions to the $O(\nu^2)$ contribution to the EOB radial potential $A(u ; \nu)$]. We indicate the expected singular
behaviors, near the lightring, of the second-self-force-order contributions to both the redshift and the EOB $A$ potential. Our results should help both in extracting
information of direct dynamical significance from ongoing second-self-force-order computations, and in parametrizing their global strong-field behaviors. We also
advocate computing second-self-force-order conservative quantities by iterating the  time-symmetric Green-function in the background spacetime.\\

{\it Dedicated to Steven Detweiler, in memoriam}
\end{abstract}
\maketitle

\section{Introduction}
In recent years, a useful synergy has developed between various approaches to the general relativistic two-body problem. The effective one-body (EOB) formalism \cite{Buonanno:1998gg,Buonanno:2000ef,Damour:2000we,Damour:2001tu} has played a special role within this synergy because it can incorporate information coming from very different ways of tackling the two-body problem, such as post-Newtonian (PN) theory, self-force (SF) theory, and numerical relativity.

The aim of the present work is to give explicit formulas exhibiting the connection between Detweiler's \cite{Detweiler:2008ft} redshift function (along circular orbits)
\beq
\label{z}
z\left(m_2\Omega; \frac{m_1}{m_2}\right)=\left( \frac{d\tau}{dt} \right)_{\mathcal L_1}^{\rm reg}\,,
\eeq
or its inverse 
\beq
\label{U}
U\left(m_2\Omega; \frac{m_1}{m_2}\right)=\left( \frac{dt}{d\tau} \right)_{\mathcal L_1}^{\rm reg}=\frac{1}{z\left(m_2\Omega; \frac{m_1}{m_2}\right)}\,,
\eeq
and the basic radial potential describing the dynamics of circular orbits in EOB theory:
\beq
\label{A}
A(u;\nu)=-g_{00}^{\rm eff}(R,m_1,m_2)\,.
\eeq

Our notation here is as follows. The two masses of  the considered  (non-spinning) binary system are $m_1$ and $m_2$, with the convention $m_1\le m_2$ (and $m_1\ll m_2$ in SF calculations).
We consider  a circular motion of orbital frequency $\Omega$. In Eqs. \eqref{z}, \eqref{U}, $d\tau$ refers to the proper time along the world line ${\mathcal L}_1$ of $m_1$, as measured in the regular, conservative part of the perturbed metric. In Eq. \eqref{A} $g_{00}^{\rm eff}$ is the time-time component of the effective EOB metric, which depends on the (Schwarzschild-like) radial coordinate $R$, while $u\equiv M/R$ (in the units $G=c=1$ we use). 
We follow the standard EOB notation
\begin{eqnarray}
M&\equiv & m_1+m_2\,,\quad \mu\equiv \frac{m_1m_2}{m_1+m_2}\,,\nonumber\\
\nu&\equiv &\frac{\mu}{M}=\frac{m_1m_2}{(m_1+m_2)^2}\,.
\end{eqnarray}

Note that while EOB theory works with symmetric functions of $m_1$ and $m_2$, SF theory considers functions of $m_2\Omega$, expanded in powers of $m_1/m_2$. Let us also introduce the notation
\begin{eqnarray}
x&\equiv &[(m_1+m_2)\Omega]^{2/3}\,,\qquad y\equiv  (m_2\Omega)^{2/3}\,,\nonumber\\
q&\equiv &\frac{m_1}{m_2}\,.
\end{eqnarray}
In terms of this notation, we can formulate the aim of the present work as follows. We wish to connect the $\nu$-expansion of the EOB $A$-potential
\beq
\label{Aexp}
A(u;\nu) =1-2u +\nu a_1(u)+\nu^2 a_2(u) +O(\nu^3)
\eeq
to the $q$-expansion (or SF-expansion) of the redshift functions, Eqs. \eqref{z}, \eqref{U},
\begin{eqnarray}
\label{zexp}
z(y;q)&=& \sqrt{1-3y}+q z_{\rm 1SF}(y)+q^2 z_{\rm 2SF}(y)+O(q^3)\nonumber\\
\label{Uexp}
U(y;q)&=&  \frac{1}{\sqrt{1-3y}}+q U_{\rm 1SF}(y)+q^2 U_{\rm 2SF}(y)+O(q^3)\,.\nonumber\\
\end{eqnarray}
The tools we shall use to connect the expansions \eqref{Aexp} and \eqref{zexp} 
are, on the one hand, the basic EOB results about the energetics of circular orbits \cite{Damour:2009sm}, and, on the other hand, the first law of binary dynamics \cite{LeTiec:2011ab}. The use of these tools at the first SF (1SF) order has led to a simple relation between the $O(\nu)$ contribution, $a_1(u)$, to the EOB  $A$ potential, Eq. \eqref{Aexp}, and the 1SF (i.e., $O(q)$) contribution, $z_{\rm 1SF}(y)$, to the redshift, namely  \cite{Barausse:2011dq}
\beq
\label{z1SF}
z_{\rm 1SF}(y)=\frac{a_1(y)}{\sqrt{1-3y}}+\frac{y(1-4y)}{1-3y}\,.
\eeq
Note in passing that the first derivation of Eq. \eqref{z1SF} proceeded via the functional link $E(x)$ between the energy of the binary system and the frequency parameter $x=[(m_1+m_2)\Omega]^{2/3}$, and, in view of the results of Ref. \cite{LeTiec:2011dp}, had to solve a first-order differential equation in $z_{\rm 1SF}$ to get the simple link  \eqref{z1SF}. A direct proof that the link between  $z_{\rm 1SF}$ and $a_1$ is {\it algebraic}, and does not involve any differentiation, has been recently given in Eq. (2.9) of Ref. \cite{Bini:2015xua}, using general properties of Legendre transforms. 
The link \eqref{z1SF} relates the two {\it functions}  $z_{\rm 1SF}(\cdot): y \to z_{\rm 1SF}(y)$ and $a_1(\cdot): u \to a_1(u)$, as defined by the expansions \eqref{zexp} and \eqref{Aexp}. Beware, in particular, that when performing SF expansions in powers of $q=m_1/m_2$, we keep both $m_2$ and $\Omega$ (and therefore $y=(m_2\Omega)^{2/3}$) fixed. In some papers (and, notably, in Refs. \cite{LeTiec:2011dp,Barausse:2011dq}), one expands $z$ in powers of $\nu$, keeping $x=[(m_1+m_2)\Omega]^{2/3}$ fixed. This changes the meaning of the expansion coefficients in
\beq
z(x,\nu)=\sqrt{1-3x}+\nu z_{(x)}^{\rm 1SF}(x) +\nu^2 z_{(x)}^{\rm 2SF}(x)+O(\nu^3)\,.
\eeq
For instance, in view of the (exact) relations 
\beq
x= (1+q)^{2/3}y\,,\qquad \nu=\frac{q}{(1+q)^2}\,,
\eeq
$z_{(x)}^{\rm 1SF}(x)$ differs from $z_{\rm 1SF}(x)$ already at first SF order:
\beq
z_{(x)}^{\rm 1SF}(x)=z_{\rm 1SF}(x) + \frac{x}{\sqrt{1-3x}}\,.
\eeq
[Here, as elsewhere, $z_{(x)}^{\rm 1SF}(x)$ and $z_{\rm 1SF}(x)$ denote the values of the functions $z_{(x)}^{\rm 1SF}(\cdot)$ and $z_{\rm 1SF}(\cdot)$ at the same, generic, argument, denoted $x$.]

The 1SF-order link \eqref{z1SF} has been quite useful for translating 1SF results on the redshift into dynamical information of relevance for binary systems \cite{Blanchet:2009sd,Blanchet:2010zd,Barausse:2011dq,LeTiec:2011dp,Akcay:2012ea,Barack:2010ny,Bini:2013zaa, Bini:2013rfa, Bini:2015bla}.

Note, in particular, that the recent derivation of the 4PN dynamics \cite{Damour:2014jta,Damour:2015isa,Damour:2016abl} (see also \cite{Bernard:2015njp}) has made a crucial use of the 1SF results of Ref. \cite{Bini:2013zaa}. Similar links have extracted useful dynamical information about more complicated binary configurations (eccentric, spinning, tidally interacting) from corresponding 1SF results \cite{Damour:2009sm,Barack:2011ed,Dolan:2013roa,Bini:2014ica,Dolan:2014pja,Bini:2014zxa,Tiec:2014lba,Bini:2015mza,Bini:2015xua,Bini:2015bfb,Akcay:2015pza,Hopper:2015icj,Bini:2016qtx,Bini:2016dvs}.

In this work, we shall consider the simplest (circular, nonspinning) binary configuration, but we shall generalize the first SF-order link \eqref{z1SF} to the second SF-order.
Indeed, after preparatory theoretical works on second-order SF (2SF) theory \cite{Pound:2009sm,Rosenthal:2006nh,Rosenthal:2006iy,Detweiler:2011tt}, there now seems to exist practical means of concretely computing 2SF effects. In particular, Ref. \cite{Pound:2014koa} has shown how to implement, and compute, Detweiler's redshift functions \eqref{z}, \eqref{U}
at the 2SF level, so as to provide a 2SF gauge-invariant measure of the {\it conservative} effects on (quasi-)circular orbits. Note that there are subtleties in the definition of the conservative dynamics at the 2SF order, which are linked to delicate infrared effects \cite{Pound:2015wva}. Some of these subtleties have been recently addressed, within the post-Minkowskian theory of Fokker actions  \cite{Damour:1995kt}, in the discussion of the nonlocal 4PN action \cite{Damour:2016abl}. We shall comment again on these subtleties in our Conclusions. Separately from these, our text will show how to transcribe a conservative 2SF redshift into a corresponding 2SF contribution to the conservative EOB Hamiltonian.

\section{The redshift in the EOB formalism}

Let us first show how one can derive an {\it exact} expression for the redshift \footnote{In this section, we reinstate a label 1 on the redshift $z_1=1/U_1$ associated with the world line ${\mathcal L}_1$ of $m_1$ (and a label 2 on the corresponding quantities associated with the mass $m_2$).} $z_1=z$, Eq. \eqref{z} of the particle $m_1$ as a function of the EOB gravitational potential $u=M/R$.  We recall that Ref. \cite{LeTiec:2011ab} (see also \cite{Blanchet:2012at}) has shown that $z_1$ and $z_2$ (along circular orbits) are related to the Hamiltonian of the binary system by
\begin{eqnarray}
\label{z1}
z_1 &=& \left[\frac{\partial}{\partial m_1} H(R,P_R,P_\phi,m_1,m_2)   \right]^{\rm circ}\nonumber\\
&=&\frac{\partial}{\partial m_1} H^{\rm circ}(P_\phi,m_1,m_2) \nonumber\\
\label{z2}
z_2 &=& \left[\frac{\partial} {\partial m_2} H(R,P_R,P_\phi,m_1,m_2)  \right]^{\rm circ}\nonumber\\
&=&\frac{\partial} {\partial m_2} H^{\rm circ}(P_\phi,m_1,m_2)\,.
\end{eqnarray}
Here $P_\phi$ is the total angular momentum of the system, which must be kept fixed during the differentiation with respect to (wrt) the masses. The superscript \lq\lq circ" indicates that one works along the sequence of circular orbits, submitted to the constraints
\beq
\label{circcons}
P_R=0\,,\qquad \frac{\partial}{\partial R} H(R,P_R,P_\phi,m_1,m_2)=0\,.
\eeq
Because of the latter constraint [and of the $O(P_R^2)$ dependence of $H$], one can also (as indicated above) evaluate $z_1$ by first imposing the constraints \eqref{circcons} to express $H$ as a function of $P_\phi$ and the masses, and then differentiating the resulting function $ H^{\rm circ}(P_\phi,m_1,m_2)$ wrt $m_1$ (keeping $P_\phi$ and $m_2$ fixed).

EOB theory expresses the Hamiltonian of the binary system in the form
\beq
\label{H}
 H(R,P_R,P_\phi,m_1,m_2)=M \sqrt{1+2\nu \left(\frac{H_{\rm eff}}{\mu}-1\right)}
\eeq
where the effective Hamiltonian $H_{\rm eff}$ reads
\begin{widetext}
\beq
\label{Heff}
H_{\rm eff}(R,P_R,P_\phi,m_1,m_2)=\sqrt{A\left(\frac{M}{R};\nu \right)\left(\mu^2+\frac{P_\phi^2}{R^2}+\frac{P_R^2}{B(M/R;\nu)}+Q(R,P_R,M,\nu)  \right)}\,.
\eeq
\end{widetext}
Here, $B$ and $Q$ are EOB potentials associated with the description of eccentric orbits. [We use the Damour-Jaranowski-Sch\"afer gauge \cite{Damour:2000we} in which $Q=O(P_R^4)$.]
When considering the energetics (and the redshift) along circular orbits one can set $P_R=0$ from the beginning so that the extra EOB potentials $B$ and $ Q$ disappear, and all results will only depend on the main EOB radial potential $A(u;\nu)$. 
More precisely, the effective potential determining the sequence of circular orbits can be taken as being
\beq
\hat H_{\rm eff}^2 =A(u;\nu)(1+p_\phi^2 u^2)\,,
\eeq
where
\beq
\hat H_{\rm eff}=\frac{ H_{\rm eff}}{\mu}\,,\quad u=\frac{M}{R}\,,\quad p_\phi=\frac{P_\phi}{\mu M}\,.
\eeq
The condition $(\partial H/\partial R)_{P_\phi}=0$ is equivalent to the condition $(\partial \hat H_{\rm eff}^2/\partial u)_{p_\phi}=0 $ and yields the circular condition
\beq
\label{pphicirc}
\left(p_\phi^2\right)^{\rm circ}=-\frac{\partial_u A}{\partial_u(u^2A)}=-\frac{\partial_u A}{2u \tilde A}\,,
\eeq
where we have defined (following \cite{Bini:2012gu})
\beq
\tilde A(u; \nu) \equiv A(u;\nu)+\frac12 u \partial_u A(u;\nu)\,.
\eeq
On the other hand, before inserting the circular solution \eqref{pphicirc}, the partial derivative of \eqref{H} wrt $P_\phi$ yields the orbital frequency
\beq
\label{Omega}
\Omega= \frac{\partial}{\partial P_\phi}  H(R,P_R,P_\phi,m_1,m_2)=\frac{MAP_\phi}{R^2 H  H_{\rm eff}}\,.
\eeq
In EOB (and PN) theory it is convenient to work with the dimensionless variables $M\Omega$, $u$, $p_\phi$, $\hat H_{\rm eff}$ and
\beq
\label{eq:210}
h=\frac{H}{M}=\sqrt{1+2\nu \left(\frac{H_{\rm eff}}{\mu}-1\right)}\,.
\eeq
In terms of these, Eq. \eqref{Omega} reads
\beq
M\Omega = \frac{ A u^2 p_\phi}{ h \hat H_{\rm eff}}\,.
\eeq
The use of the circular condition \eqref{pphicirc} allows one to express all physical quantities, along the sequence of circular orbits, as explicit functions of $u$. In particular, we have
\beq
\label{eq:212}
\hat H_{\rm eff}^{\rm circ}=\frac{A}{\sqrt{\tilde A}}\,,\qquad h^{\rm circ}=\sqrt{1+2\nu \left( \frac{A}{\sqrt{\tilde A}} -1 \right)}
\eeq
and
\begin{eqnarray}
\label{MOmegau}
M\Omega^{\rm circ}&=&\frac{u^{3/2}}{h^{\rm circ}}\sqrt{-\frac12 \partial_u A}\nonumber\\
&=& u^{3/2} \sqrt{\frac{-\frac12 \partial_u A}{1+2\nu \left( \frac{A}{\sqrt{\tilde A}} -1 \right)}}\,.
\end{eqnarray}
We can also straightforwardly evaluate $z_1$ and $z_2$ from Eqs. \eqref{z1}. In doing so, we must remember that $P_\phi\equiv m_1m_2p_\phi$ must be kept fixed during the differentiation wrt the masses. Alternatively, we can compute the total derivative of $H$ along the sequence of circular motions (parametrized, say, by $u$, $m_1$ and $m_2$) using the identity \cite{LeTiec:2011ab}
\beq
dH=\Omega dP_\phi +z_1 dm_1+z_2dm_2\,.
\eeq
After some simplifications, one finds that $z_1$ and $z_2$ can be expressed by compact, explicit functions of $u$, namely
\begin{eqnarray}
\label{z1u}
z_1(u, \nu)&=& \frac{1}{h^{\rm circ}}\left[X_1 +X_2 \sqrt{\tilde A}+\frac{\nu}{2}X_2 X_{21}\frac{\partial_\nu A}{\sqrt{\tilde A}}  \right] \, \\
\label{z2u}
z_2(u, \nu)&=& \frac{1}{h^{\rm circ}}\left[X_2 +X_1 \sqrt{\tilde A}+\frac{\nu}{2}X_1 X_{12}\frac{\partial_\nu A}{\sqrt{\tilde A}}  \right]\,,\nonumber\\
\end{eqnarray}
where $X_1=m_1/M$, $X_2=m_2/M$, $X_{12}=X_1-X_2=-X_{21}$, and where all other variables are considered as functions of $u$ and $\nu$
(with $\partial_\nu A \equiv \partial A(u; \nu)/ \partial \nu$).
Note that $X_1+X_2=1$, $\nu=X_1X_2$ and that, under the convention $m_1\le m_2$, one has 
\beq
X_1=\frac12 (1-\sqrt{1-4\nu})\,,\qquad X_2=\frac12 (1+\sqrt{1-4\nu})\,.
\eeq
In order to compare the result \eqref{z1u} to SF calculations, in which $z=z_1$ is considered as a function of $\Omega$, see Eq. \eqref{zexp}, we need to invert the function $u\to \Omega$ defined by Eq. \eqref{MOmegau}. This is straightforward to do, if one expands in powers of $\nu$ or $q$. Indeed, as $A(u;\nu)=1-2u+O(\nu)$, we see that Eq. \eqref{MOmegau} is of the form $M\Omega=u^{3/2}(1+O(\nu))$.
When $\nu\to0$ (i.e., $q=m_1/m_2\to 0$) we recover Kepler's law (in a Schwarzschild spacetime). 

When working with the dimensionless frequency parameter $x=(M\Omega)^{2/3}$, Eq. \eqref{MOmegau} reads
\beq
\label{xu}
x=u \left( \frac{-\frac12 \partial_u A}{1+2\nu \left(\frac{A}{\sqrt{\tilde A}}-1\right)} \right)^{1/3}\,.
\eeq
Inserting in Eq. \eqref{xu} the $\nu$-expansion of $A$, Eq. \eqref{Aexp}, we can straightforwardly compute the $\nu$-expansion of $x=u(1+O(\nu))$ as a function of $u$, namely
\begin{eqnarray}
x&=&u +\nu U_1(u; a_1'(u))\nonumber\\
&+& \nu^2 U_2(u;a_1(u),a_1'(u),a_2'(u))+O(\nu^2)\,,
\end{eqnarray}
where
\begin{eqnarray}
 U_1  
&=&  -\frac16 u \left[ a_1'(u)-4 \left(1-\frac{1-2u}{\sqrt{1-3u}}  \right)\right]\nonumber\\
 U_2 
&=&  -\frac13 \frac{u (1-4 u)}{ (1-3 u)^{3/2}} a_1(u)
-\frac16  u a_2'(u)
-\frac{1}{36}u [a_1'(u)]^2\nonumber\\
&&+\left(\frac{ u (1-2 u) (2-3 u)}{18(1-3 u)^{3/2}}-\frac19  u\right)a_1'(u)\nonumber\\
&&
-\frac{16 u (1-2 u)}{9 (1-3 u)^{1/2}}+\frac{8u(2-7 u+4 u^2)}{9(1-3 u)}  \,.
\end{eqnarray}
Inverting this functional link then yields
\begin{eqnarray}
\label{u_di_x}
u&=&x -\nu U_1(x; a_1'(x)) \nonumber\\
&&+\nu^2 V_2(x;a_1(x),a_1'(x),a_2'(x))+O(\nu^2)\,,
\end{eqnarray}
where
\begin{eqnarray}
V_2 
&=& U_1(x; a_1'(x)) \left(\frac{d}{dx}U_1(x; a_1'(x))\right)\nonumber\\
&&-U_2(x;a_1(x),a_1'(x),a_2'(x))\,.
\end{eqnarray}
Inserting this expansion in Eq. \eqref{z1u} yields the $\nu$-expansion of the function $z_{(x)}:x\to z$, namely
\begin{widetext}
\beq \label{z1x}
z_1(u(x, \nu), \nu) =z_{(x)}(x, \nu)=\sqrt{1-3x}+\nu z_{(x)}^{\rm 1SF}(x)+\nu^2 z_{(x)}^{\rm 2SF}(x)+O(\nu^3)\,,
\eeq
with
\begin{eqnarray} \label{z1xexp}
 z_{(x)}^{\rm 1SF}(x)&=&\frac{1}{\sqrt{1-3x}}\left[ a_{1}(x)+x \left(1+\frac{1-4x}{\sqrt{1-3x}}  \right) \right] \nonumber\\
 z_{(x)}^{\rm 2SF}(x)&=& \frac{x}{18}\left[\frac{3+18x-234 x^2 +432x^3}{(1-3x)^{5/2}} +\frac{12(2-13x+24x^2)}{(1-3x)^2} \right] +\frac{3}{2\sqrt{1-3x}}a_{2}(x) +
\frac{x}{8\sqrt{1-3x}}[ a_{1}'(x)]^2\nonumber\\
&& -\frac{3}{8(1-3x)^{3/2}}[ a_{1}(x)]^2-\frac{x}{\sqrt{1-3x}}\left(\frac23 -\frac{1-2x}{\sqrt{1-3x}}  \right)a_{1}'(x)\nonumber\\
&& -\left[\frac32 \frac{(1-2x)(1-4x)}{(1-3x)^2}
+\frac{1-2x}{(1-3x)^{3/2}}\right]a_{1}(x)\,.
\end{eqnarray}
\end{widetext}
Similarly, inserting the $\nu$-expansion of the function $u_{(x)}:x\to
u$, Eq. \eqref{u_di_x}, in the expression of the EOB Hamiltonian in terms of
$u$, Eqs. \eqref{eq:210}, \eqref{eq:212}, yields the $\nu$-expansion of the fractional
binding energy, $\hat E=(H-M)/\mu$, expressed as a function of the
frequency parameter $x$. Its structure is more complicated than that of
the function $z_{(x)}$ because it involves a derivative of  $a_1$
already at the $O(\nu)$ order \cite{Barausse:2011dq,Akcay:2012ea}. 
At the $O(\nu^2)$ order it is quadratic in $a_1$ and its first and second derivatives.
It reads
\beq
\hat E(x; \nu)=e_ 0(x)+\nu e_{1}(x)+\nu^2 e_{2}(x)+O(\nu^3)\,,
\eeq
with
\beq
e_0(x)=\frac{1-2x}{\sqrt{1-3x}}-1\,,
\eeq
and  the $O(\nu)$ contribution given by (consistently with \cite{Barausse:2011dq,Akcay:2012ea}) 
\begin{eqnarray}
e_{1}(x)&=& -\frac13 \frac{x}{\sqrt{1-3x}}a_1'(x)+\frac{1}{2}\frac{1-4x}{(1-3x)^{3/2}}a_1(x)\nonumber\\
&&-e_0(x)\left[ \frac12 e_0(x)+\frac{x}{3}\frac{1-6x}{(1-3x)^{3/2}} \right]\,.
\end{eqnarray}
The expression for the $O(\nu^2)$ contribution $e_{2}(x)$ is much more involved and can be decomposed as
\begin{eqnarray}
e_{2}(x)&=& e_{2(0)}(x)+ e_{2(a_1)}(x)+ e_{2(a_2)}(x)\,,
\end{eqnarray}
with $e_{2(a_1)}(x)=e_{2(a_1)^2}(x)+e_{2(a_1)^1}(x)$ and
\begin{widetext}
\begin{eqnarray}
e_{2(0)}(x)&=& \frac{(1-2\sqrt{1-3x})(\sqrt{1-3x}-1)^3}{486 (1-3x)^{7/2}}\left[3(7x-2)(18x-7)\sqrt{1-3x}-549x^2 +285x-39  \right]\\
e_{2(a_1)^2}(x)&=& \frac{x(1-6x)}{72(1-3x)^{3/2}}[a_1'(x)]^2+\frac{x}{6\sqrt{1-3x}}\left[ \frac{a_1}{(1-3x)}-\frac{x}{3}a_1''(x) \right] a_1'(x)-\frac{(1-6x)}{8 (1-3x)^{5/2}} [a_1(x)]^2\\
e_{2(a_1)^1}(x)&=& -\frac{x(1-\sqrt{1-3x})}{27(1-3x)^2}[(7-30x)\sqrt{1-3x}-5+24x]a_1'(x)-\frac{2x^2}{27(1-3x)}(1-\sqrt{1-3x})(1-2\sqrt{1-3x})a_1''\nonumber\\
&& -\frac{(1-\sqrt{1-3x})(1-2\sqrt{1-3x})}{18(1-3x)^3}[2-14x+36x^2 -(1-6x)\sqrt{1-3x}]a_1(x)\\
 e_{2(a_2)}(x)&=& \frac{1-4x}{2(1-3x)^{3/2}}a_2(x)- \frac{x}{3\sqrt{1-3x}}a_2'(x)\,.
\end{eqnarray}
\end{widetext}

As a last step, to be closer to what is actually computed in SF theory, we must replace $x$ by $(1+q)^{2/3}y$ and $\nu$ by $q/(1+q)^2$ in order to derive the $q$-expansion of $z(y,q)$, Eq. \eqref{zexp}. In doing this transformation we need expansions of the type
\begin{eqnarray}
f[(1+q)^{2/3}y]&=& f(y)+\frac23y q f'(y)\\
&+&\frac19 y q^2\left[- f'(y) +2y  f''(y) \right]+O(q^3)\,.\nonumber
\end{eqnarray}
Our final result for the coefficients in the SF-expansion \eqref{zexp} read
\begin{widetext}
\begin{eqnarray}
\label{z1_y_fin}
 z_{\rm 1SF}(y)&=& \frac{a_1(y)}{\sqrt{1-3y}}+\frac{y(1-4y)}{1-3y}\, , \nonumber\\
 z_{\rm 2SF}(y)&=& \frac32 \frac{a_{2}(y)}{\sqrt{1-3y}}+\frac18 \frac{y}{\sqrt{1-3y}}[ a_{1}'(y)]^2+\frac{y(1-2y)}{(1-3y)} a_{1}'(y)-\frac{3}{8}\frac{[ a_{1}(y)]^2}{(1-3y)^{3/2}}\nonumber\\
&& -\left(\frac{3(1-2y)(1-4y)}{2(1-3y)^2} +\frac{3}{\sqrt{1-3y}}\right) a_{1}(y)-\frac12 \frac{y(1-2y)}{(1-3y)^{5/2}}(2-13y +24 y^2)\,.
\end{eqnarray}
\end{widetext}

As the function $a_1(u)$ is accurately known (numerically \cite{Akcay:2012ea} and analytically \cite{Akcay:2012ea,Bini:2015bla,Kavanagh:2015lva}), Eq. \eqref{z1_y_fin} shows that one can algebraically compute the function $a_2(u)$ from the knowledge of $z_{\rm 2SF}(y)$.

Let us complete these results by giving the corresponding SF-expansions of the inverse redshift $U(y,q)\equiv 1/z(y,q)$, as well as the SF expansions of the ratios
\begin{eqnarray}
\label{widehatz1andU_y_fin}
\widehat z(y;q)&=&\frac{z_(y;q)}{\sqrt{1-3y}}\,,\nonumber\\
\widehat U(y;q)&=&\sqrt{1-3y}\,  U(y;q)=\frac{1}{\widehat z(y;q)}\,.
\end{eqnarray}
They read
\begin{eqnarray}
\label{avatars}
U(y)&=& \frac{1}{\sqrt{1-3y}}+q U_{\rm 1SF}(y)+q^2 U_{\rm 2SF}(y)+O(q^3)\nonumber\\
\widehat z(y;q) &=& 1+q \widehat z_{\rm 1SF}(y)+q^2 \widehat z_{\rm 2SF}(y)+O(q^3)\nonumber\\
\widehat U(y;q) &=& 1+q \widehat U_{\rm 1SF}(y)+q^2 \widehat U_{\rm 2SF}(y)+O(q^3)\,,
\end{eqnarray}
where
\begin{eqnarray}
U_{\rm 1SF}(y)&=&-\frac{z_{\rm 1SF}(y)}{1-3y}\nonumber\\
U_{\rm 2SF}(y)&=& -\frac{z_{\rm 2SF}(y)}{(1-3y)}+\frac{[z_{\rm 1SF}(y)]^2}{(1-3y)^{3/2}}  \nonumber\\
\widehat z_{\rm 1SF}(y)&=& \frac{z_{\rm 1SF}(y)}{\sqrt{1-3y}}\nonumber\\
\widehat z_{\rm 2SF}(y)&=&\frac{z_{\rm 2SF}(y)}{\sqrt{1-3y}}\nonumber\\
\widehat U_{\rm 1SF}(y)&=& -\frac{z_{\rm 1SF}(y)}{\sqrt{1-3y}}\nonumber\\
\widehat U_{\rm 2SF}(y)&=& -\frac{z_{\rm 2SF}(y)}{\sqrt{1-3y}}+\frac{[z_{\rm 1SF}(y)]^2}{(1-3y)} \,.
\end{eqnarray}

\section{PN expansion of the second-order redshift}

The PN-expansion of the Hamiltonian of a binary system is currently fully known through the 4PN level \cite{Damour:2014jta,Damour:2015isa,Damour:2016abl}. In addition, some higher PN contributions are known. This is the case for the logarithmic contributions at the 5PN level, see \cite{Blanchet:2010zd} (with corrections given in \cite{LeTiec:2011ab}), \cite{Damourlogs} (whose derivation was given in \cite{Damour:2015isa}), and \cite{Barausse:2011dq}. In particular the 4PN and 5PN logarithmic contributions to the EOB $A$ potential are (see, e.g., Eq. (9.14a) of Ref.  \cite{Damour:2015isa})
\beq
\label{Alog}
A^{\ln{}}(u;\nu)=\frac{64}{5}\nu u^5 \ln u+\left( -\frac{7004}{105}\nu -\frac{144}{5}\nu^2 \right) u^6 \ln u\,.
\eeq
As we see, while the 4PN-level logarithmic contribution to $A$ is linear in $\nu$, the 5PN-level logarithmic contribution is quadratic in $\nu$.
Let us recall in this respect that remarkable cancellations take place in the $\nu$-dependence of the EOB $A$ potential. Indeed, while, for instance, the PN expansion of the fractional binding energy, $\hat E=(H-M)/\mu$, expressed as a function of the frequency parameter $x$, has a nonlinear dependence on $\nu$ which starts already at the 2PN level, say (without indicating the logarithmic running of the 4PN terms) \cite{Damour:2014jta}
\begin{widetext}
\begin{eqnarray}
\hat E(x; \nu)&=& -\frac12 x \left[ 1+(e_1+e_1'\nu)x
+(e_2+e_2'\nu+e_2''\nu^2) x^2
+(e_3+e_3'\nu+e_3''\nu^2+e_3'''\nu^3 ) x^3 \right. \nonumber\\
&& \left.
+(e_4+e_4'\nu+e_4''\nu^2+e_4'''\nu^3 +e_4'''' \nu^4 ) x^4+\ldots\right]\,,
\end{eqnarray}
\end{widetext}
the EOB potential $A$ stays linear in $\nu$ through the 3PN level \cite{Damour:2000we}, and features (only) a $O(\nu^2)$ nonlinearity at the 4PN level \cite{Bini:2013zaa} \footnote{We recall that the nPN level in $A(u)$ corresponds to a term $\propto u^{n+1}$.}
\beq
A(u;\nu)=1-2u+2\nu u^3 +\nu a_4 u^4+(\nu a_5' +\nu^2 a_5'')u^5 +\ldots\,.
\eeq
In view of Eq. \eqref{z1_y_fin}, this immediately indicates that the first new information contained in the 2SF redshift $z_{2\rm SF}(y)$ will start with the nonlogarithmic 5PN contribution, i.e., $z_{2\rm SF}^{\rm new}(y)\propto y^6$ (see below for its explicit parametrization). [We do not discuss here the PN expansion of the first-order SF terms $a_1(u)$ or $z_{1\rm SF}(y)$ which are analytically known to high PN orders  \cite{Bini:2015bla,Kavanagh:2015lva}, and numerically known up to $u=\frac13 $ \cite{Akcay:2012ea}; see Eq. \eqref{a165pn} below, and Appendix A.]

 Let us also mention that Ref. \cite{Bini:2013rfa} has argued that the first
 PN contribution to $A(u ; \nu)$ that is {\it cubic} in $\nu$ will start
 at the 6PN order, i.e.  that $a_3(u) = O(u^7)$, and that the first PN
 contribution that is {\it quartic} in $\nu$ will start at the 8PN order, etc. 
 This indicates that
 the knowledge of  the 2SF redshift (which gives, in principle, access
 to the function $a_2(u)$)  gives also access to many low-order
 contributions in the $\nu$ expansion of the function $\hat E(x; \nu)$.

Some information is known about the 2SF  contributions to the half-integer 5.5PN level. Indeed, while the 1SF derivations of the 5.5PN-level contribution to the redshift \cite{Shah:2013uya,Bini:2013rfa} do not give any information about the 2SF level, the corresponding PN-based derivations \cite{Blanchet:2013txa,Damour:2015isa}, especially that of the latter reference which directly computed the 5.5PN-level contribution to the $A$ potential  (see Eq. (9.32) in \cite{Damour:2015isa}), show that it is linear in $\nu$, namely
\beq
\label{A55}
A^{5.5\rm PN}(u;\nu)=\frac{13696}{525}\pi \nu u^{13/2}\,.
\eeq

On the other hand, it is not clear to us whether the PN derivation of the 6.5PN nearzone metric, and associated redshift, in Ref. \cite{Blanchet:2014bza} was limited to the contribution that is linear in $\nu$, or whether it kept the terms of order $O(\nu^2)$. It would be useful that the authors of Ref. \cite{Blanchet:2014bza} re-examine their proof and state their result in terms of the 6.5PN contribution to the EOB $A$ potential to know what is the value of the coefficient $a'_{7.5}$ of $\nu^2$ in
\beq
\label{A65}
A^{6.5\rm PN}(u;\nu)=\left(-\frac{512501}{3675}\pi \nu+a'_{7.5}\nu^2 \right)u^{15/2}\,.
\eeq
Starting from the known terms in the PN expansion of the EOB $A$ potential, i.e. the 4PN \cite{Damour:2014jta,Damour:2015isa}, 5PN logs, Eq. \eqref{Alog}, 5.5PN \cite{Damour:2015isa,Blanchet:2013txa}, together with all the terms that are known to first order in $\nu$ 
\cite{Bini:2013zaa,Shah:2013uya,Bini:2013rfa,Bini:2014nfa,Johnson-McDaniel:2015vva,Bini:2015bla,Blanchet:2014bza}, and parametrizing the $O(\nu^2)$ terms that are still unknown, we can write the PN expansion of the contributions $a_1(u)$ and $a_2(u)$ in Eq. \eqref{Aexp} as
\begin{eqnarray}  
\label{a165pn}
a_1(u)&=& 2u^3 +\left(\frac{94}{3}-\frac{41}{32}\pi^2  \right) u^4\nonumber\\
&+ & \left(-\frac{4237}{60}+\frac{2275}{512}\pi^2+\frac{128}{5}\gamma+\frac{256}{5}\ln(2)\right. \nonumber\\
&& \left. +\frac{64}{5}\ln(u )\right ) u^5 +O(u^6)\,,
\end{eqnarray}
(see Appendix A for additional terms) 
and
\begin{eqnarray}
a_2(u)&=& \left(-\frac{221}{6}+\frac{41}{32}\pi^2\right)u^5+\left(-\frac{144}{5}\ln(x)+a_6' \right) u^6\nonumber\\
&& +\, 0\, u^{13/2}+(a_7'{}^{\ln{}}\ln u +a_7')u^7 +a_{7.5}'u^{7.5}+\ldots\nonumber\\
\end{eqnarray}
Inserting these expressions in Eqs. \eqref{z1_y_fin} and \eqref{widehatz1andU_y_fin} we get the following PN-expanded expressions for the various ways of defining the 2SF redshift contribution:
\begin{eqnarray}
\label{splitK_unK}
z_{2\rm SF}^{\rm PN}(y)&=& z_{2\rm SF}^{\rm known}(y)+z_{2\rm SF}^{a_2\rm -unknown}(y)\nonumber\\
\widehat z_{2\rm SF}^{\rm PN}(y)&=& \widehat z_{2\rm SF}^{\rm known}(y)+\widehat z_{2\rm SF}^{a_2\rm -unknown}(y)\nonumber\\
U_{2\rm SF}^{\rm PN}(y)&=& U_{2\rm SF}^{\rm known}(y)+U_{2\rm SF}^{a_2\rm -unknown}(y)\nonumber\\
\widehat U_{2\rm SF}^{\rm PN}(y)&=& \widehat U_{2\rm SF}^{\rm known}(y)+\widehat U_{2\rm SF}^{a_2\rm -unknown}(y)\,,
\end{eqnarray} 
namely
\begin{widetext}
\begin{eqnarray}
\label{z2K}
z_{2\rm SF}^{\rm known}(y)&=&-y+y^2-\frac{29}{8}y^3+\left(-\frac{74}{3}+\frac{41}{64}\pi^2\right) y^4+\left(\frac{64}{5}\gamma+\frac{128}{5}\ln(2)+\frac{4899}{1024}\pi^2+\frac{32}{5}\ln(y)-\frac{75107}{640}  \right) y^5\nonumber\\
&&+\left(-\frac{6556}{35}\gamma-\frac{14972}{35}\ln(2)+\frac{729}{14}\ln(3)+\frac{232221}{2048}\pi^2-\frac{958}{7}\ln(y)-\frac{66534539}{67200}  \right) y^6 \nonumber\\
&& +\frac{27392}{525}\pi y^{13/2}\nonumber\\ 
&& 
+\left(\frac{351398}{2835}\gamma+\frac{3312926}{2835}\ln(2)-\frac{21627}{28}\ln(3)+\frac{3018779419}{3538944}\pi^2-\frac{12283021}{524288}\pi^4\right. \nonumber\\
&& \left.-\frac{8009}{2835}\ln(y)-\frac{50685282659}{14515200}\right) y^7\nonumber\\
&&  -\frac{1345759}{3675}\pi   y^{15/2}\nonumber\\
\label{z2unK}
z_{2\rm SF}^{a_2\rm -unknown}(y)&=& \frac32 a'_6 y^6  
+\left(\frac94 a'_6+\frac32 a'_7+\frac32 a_7'{}^{\ln{}} \ln(y)\right) y^7+\frac32 a'_{7.5}y^{15/2}\,,
\end{eqnarray}
\begin{eqnarray}
\widehat z_{2\rm SF}^{\rm known}(y)&=&-y-\frac12 y^2-\frac{11}{2}y^3+\left(-\frac{211}{6}+\frac{41}{64}\pi^2\right) y^4+\left(-\frac{1803}{10}+\frac{5883}{1024}\pi^2+\frac{64}{5}\gamma \right. \nonumber \\
&& \left. +\frac{128}{5}\ln(2)+\frac{32}{5}\ln(y)\right) y^5\nonumber\\
&& +\left(-\frac{2766997}{2100}-\frac{5884}{35}\gamma-\frac{13628}{35}\ln(2)+\frac{125673}{1024}\pi^2-\frac{4454}{35}\ln(y)+\frac{729}{14}\ln(3)\right) y^6\nonumber\\
&& +\frac{27392}{525}\pi y^{13/2}
 \nonumber\\
&& +\left(-\frac{1046211847}{181440}+\frac{3696967147}{3538944}\pi^2-\frac{12283021}{524288}\pi^4-\frac{528758}{2835}\ln(y)\right. \nonumber\\
&&\left. +\frac{248396}{405}\ln(2)-\frac{322684}{2835}\gamma-\frac{4860}{7}\ln(3)\right) y^7 \nonumber\\
&&  -\frac{1058143}{3675}\pi y^{15/2}\nonumber\\
\widehat z_{2\rm SF}^{a_2-\rm unknown}(y)&=&  \frac32 a_6' y^6 +\left(\frac92 a_6'+\frac32 a_7'+\frac32 a_7'{}^{\ln{}}\ln(y)\right) y^7+\frac32 a_{7.5}'y^{15/2}\,,
\end{eqnarray}

\begin{eqnarray}
U_{2\rm SF}^{\rm known}(y)&=& y+3 y^2+\frac{97}{8}y^3+\left(\frac{725}{12}-\frac{41}{64}\pi^2\right) y^4+\left(\frac{674801}{1920}-\frac{9491}{1024}\pi^2-\frac{64}{5}\gamma-\frac{128}{5}\ln(2)-\frac{32}{5}\ln(y)\right) y^5\nonumber\\
&& + \left(\frac{7004}{35}\gamma+\frac{15868}{35}\ln(2)-\frac{729}{14}\ln(3)-\frac{281463}{2048}\pi^2 +\frac{5014}{35}\ln(y)  +\frac{133591739}{67200}\right) y^6\nonumber\\
&&-\frac{27392}{525}\pi y^{13/2}\nonumber\\
&&+\left( \frac{115955210999}{14515200}-\frac{3892697563}{3538944}\pi^2+\frac{12283021}{524288}\pi^4+\frac{993347}{2835}\ln(y)\right. \nonumber\\
&& \left.-\frac{590138}{2835}\ln(2)+\frac{884446}{2835}\gamma+\frac{19197}{28}\ln(3)\right) y^7\nonumber\\
&& + \frac{320757}{1225}\pi  y^{15/2}\nonumber\\
U_{2\rm SF}^{a_2-\rm unknown}(y)&=&   -\frac32 a_6' y^6 +\left(-\frac{27}{4} a_6' -\frac32  a_7'  -\frac32  a_7'{}^{\ln{}} \ln(y)\right) y^7   -\frac32  a_{7.5}'  y^{15/2}\,,
\end{eqnarray}

\begin{eqnarray}
\widehat U_{2\rm SF}^{\rm known}(y)&=& y+\frac32 y^2+\frac{13}{2}y^3+\left(\frac{223}{6}-\frac{41}{64}\pi^2\right) y^4
+\left(\frac{7169}{30}-\frac{8507}{1024}\pi^2-\frac{64}{5}\gamma-\frac{128}{5}\ln(2)-\frac{32}{5}\ln(y)\right) y^5\nonumber\\
&& +\left(\frac{2848057}{2100}+\frac{7676}{35}\gamma+\frac{17212}{35}\ln(2)-\frac{125757}{1024}\pi^2+\frac{1070}{7}\ln(y)-\frac{729}{14}\ln(3)\right) y^6\nonumber\\
&& -\frac{27392}{525}\pi y^{13/2}\nonumber\\
&& +\left(\frac{4024326563}{907200}-\frac{3122418667}{3538944}\pi^2+\frac{12283021}{524288}\pi^4+\frac{57794}{405}\ln(y)-\frac{2436452}{2835}\ln(2)\right. \nonumber\\
&& \left. +\frac{10612}{405}\gamma+\frac{5346}{7}\ln(3)\right) y^7\nonumber\\
&& + \frac{416629}{1225}\pi  y^{15/2} \nonumber\\
\widehat U_{2\rm SF}^{a_2-\rm unknown}(y)&=& -\frac32 a_6' y^6+\left(-\frac92 a_6' -\frac32 a_7'  -\frac32 a_7'{}^{\ln{}}\ln(y)\right) y^7 -\frac32 a_{7.5}'y^{15/2} \,.
\end{eqnarray} 
\end{widetext}

Here we decomposed the 2SF contributions into their analytically known parts (coming from $a_1(y)$, the analytically known part of $a_2(u)$, and the 
last, explicit, term in Eq. \eqref{z1_y_fin}), and the parts coming from the  analytically unknown part of $a_2(u)$.

We have checked that the 4PN contribution to $z_{2\rm SF}^{a_2-\rm known}(y)$ written above (as well as the full 4PN contribution to $z_{(x)}(x)$ defined by Eqs. \eqref{z1x}, \eqref{z1xexp} above) is consistent with the 4PN expansion of $z_{(x)}(x)$ derived in Ref. \cite{Tiec:2014lba} from the 4PN results of \cite{Jaranowski:2013lca,Bini:2013zaa}.

\section{Expected lightring behaviour at the 2SF level}

Ref. \cite{Akcay:2012ea} discovered that, at the 1SF level, several functions of dynamical significance had a singular behaviour at the lightring (LR), i.e., when $u\to 1/3$ or $y\to 1/3$. In particular, the ratios $\widehat U_{1\rm SF}(y)=\sqrt{1-3y}\, U_{1\rm SF}(y)$ and $\widehat z_{1\rm SF}(y)=z_{1\rm SF}(y)/\sqrt{1-3y}$ behave as
\beq
\widehat U_{1\rm SF}(y)=-\widehat z_{1\rm SF}(y)=\frac12 h_{uu}^{\rm 1SF\, reg}\propto E^3(y)\,,
\eeq
where $ h_{uu}^{\rm 1SF}=h_{\mu \nu}^{1\rm SF} u^\mu u^\nu$ ($u^\mu\equiv dx^\mu/d\tau$), and, where we introduced the notation
\beq
E(y)=\frac{1-2y}{\sqrt{1-3y}}
\eeq
for the 1SF specific energy of a test particle in a Schwarzschild spacetime. Near the LR, i.e., as $y\to \frac13^-$, $E(y)\to +\infty$.
As explained in \cite{Akcay:2012ea}, this result is (essentially) deriving from the fact that the 1SF metric perturbation $q h_{\mu \nu}^{\rm 1SF}(x)$ (at a generic field point) is sourced not only by the mass $m_1$, but, more precisely, by the energy $m_1 E$ of the particle 1. Then, the fact that $u^0=-g^{00}E$ explains why 
$h_{\mu \nu}u^\mu u^\nu$ blows up like the cube of \footnote{In this asymptotic estimate, and the ones below, we could everywhere replace $E(y)$ by
its LR-singular factor $1/\sqrt{1-3y}$.} $E$.

Pound \cite{Pound:2014koa} has derived several expressions for $\widehat U(y)$ at the 2SF accuracy. It seems that his Eq. (101)
 is the most relevant here. It reads (using his notation)
\begin{eqnarray}
\label{Pound}
\widehat U(y) &=& 1+\frac12 \widehat h_{u_0 u_0}^{R1}+\frac12 \widehat h_{u_0 u_0}^{R2}+\frac38 \left(\frac12 \widehat h_{u_0 u_0}^{R1}\right)^2\nonumber\\
&& -\frac{r_0-3m_2}{24m_2}
\left[u_0^\mu u_0^\nu \left(r\frac{\partial}{\partial r}  \widehat h_{\mu \nu}^{R1}  \right)_{r=r_0}  \right]^2\,.
\end{eqnarray}
Here, $\widehat h_{\mu\nu}^{R1}$ and $\widehat h_{\mu\nu}^{R2}$ are (respectively) precisely defined versions of the regularized 1SF and 2SF 
metric perturbations (for use in a specific 2SF scheme). We therefore expect that $\widehat h_{\mu\nu}^{R1}$ will be proportional to $qE(y)$ and $\widehat h_{\mu\nu}^{R2}$ (whose source is quadratic in 
$\widehat h_{\mu\nu}^{R1}$) to $qE^2(y)$. We then expect that the four metric-dependent contributions on the rhs of Eq. \eqref{Pound} will essentially behave (near the LR) as
\beq
\widehat U(y)\sim 1+qE^3(y)+q^2E^4(y)+q^2E^6(y)+q^2E^4(y)\,,
\eeq
so that the dominant behavior near the LR will be
\beq
\label{ULR}
\widehat U(y) \sim 1+qE^3(y)+q^2E^6(y) \,,
\eeq
as well as
\beq
\label{ZLR}
\widehat z(y)\sim 1+qE^3(y)+q^2E^6(y) \,.
\eeq
Let us note in passing that the LR behaviors \eqref{ULR} and \eqref{ZLR} are consistent with the conclusion of Section VII B of Ref. \cite{Akcay:2012ea} that the condition for the numerical validity of the SF expansion as one approaches the LR is $qE^3\ll 1$ (see Eq. (112) in Ref. \cite{Akcay:2012ea}). It would be interesting to probe the LR behavior of $\widehat U_{2\rm SF}(y)$ and $\widehat z_{2 \rm SF}(y)$ and confirm the expected behavior $\widehat U_{2\rm SF}(y)\sim \widehat z_{2 \rm SF}(y)\sim E^6(y)$.

Assuming this LR behavior, let us now turn to our EOB expressions for $\widehat U_{2\rm SF}$ and $\widehat z_{2 \rm SF}$ in terms of the EOB potentials $a_1(u)$ and $a_2(u)$. We first recall that \cite{Akcay:2012ea}  found the LR behavior 
\beq
\label{a1LR}
a_1(u) \sim E(u)\,,
\eeq
consistently with $\widehat z_{1 \rm SF} \sim E^3 $ and the link
\begin{eqnarray}
a_1(y)&=&\sqrt{1-3y}z_{1\rm SF}(y)-\frac{y(1-4y)}{\sqrt{1-3y}}\nonumber\\
&=&(1-3y)\widehat z_{1 \rm SF}-\frac{y(1-4y)}{\sqrt{1-3y}}\,.
\end{eqnarray}
We can correspondingly rewrite the second equation \eqref{z1_y_fin} in the form
\beq
\label{a2}
a_2(y)=a_2^{\rm known}(y)+a_2^{\rm unknown}(y)
\eeq
where \footnote{Here, ``known" and ``unknown" have different meanings than above.} 
\begin{widetext}
\begin{eqnarray}
\label{a2k}
a_2^{\rm known}(y)&=&  -\frac{1}{12} y[ a_{1}'(y)]^2-\frac23 \frac{y(1-2y)}{\sqrt{1-3y}} a_{1}'(y)+\frac14 \frac{[ a_{1}(y)]^2}{(1-3y)}  \nonumber\\
&&  +\left( 2+\frac{(1-2y)(1-4y)}{(1-3y)^{3/2}}\right) a_{1}(y)+\frac13 \frac{y(1-2y)}{(1-3y)^{2}}(2-13y +24 y^2) \nonumber\\
\label{a2u}
a_2^{\rm unknown}(y)&=& \frac23 (1-3y)\widehat z_{2 \rm SF}(y)\,.
\end{eqnarray}
\end{widetext}

If we insert in Eq. \eqref{a2u} the estimates $\widehat z_{2 \rm SF}\sim E^6 $, $1-3y \sim E^{-2}$, we find that the unknown, $\widehat z_{2 \rm SF}$-related, contribution to $a_2$ is expected to behave as $E^4$ near the LR. By contrast, using also the estimates $a_1(y)  \sim E(y)$ and $a_1'(y)\sim E^3(y)$, we see that the various contributions to $a_2^{\rm known}(y)$ (rhs of Eq. \eqref{a2k}) respectively behave, near the LR, as $E^6$, $E^4$, $E^4$, $E^4$ and $E^4$.
We therefore conclude that, near the LR (as $E\to \infty$) we have
\beq
a_2(y)=-\frac{1}{12}y (a_1'(y))^2+O(E^4)\,.
\eeq
In particular, as the LR behavior of $a_1(y)$ is \cite{Akcay:2012ea}
\beq
a_1(y) \simeq \frac{\zeta}{4}\frac{1}{\sqrt{1-3y}}\,,
\eeq
where the numerical value of $\zeta$ is \cite{Akcay:2012ea,Bini:2014zxa}
\beq
\zeta \approx 1.0055(5)\,,
\eeq
we conclude that the leading-order singularity of $a_2(y)$ at the LR is 
\beq
\label{a2LR}
a_2(y) \simeq -\frac{\zeta^2}{256}\frac{1}{(1-3y)^3}\,.
\eeq
Note that Eq. \eqref{a2LR} predicts that $a_2(y)$ will tend to $-\infty$  as $y\to (1/3)^-$.
[A similar prediction was made at the end of Section VII in \cite{Akcay:2012ea}, with, however, an expected milder LR singularity $\propto (1-3y)^{-2}$.]

On the other hand, the lowest-order PN contribution to $a_2(y)$ (which comes from the 4PN level) is also negative, namely
\beq
\label{a24PN}
a_2^{4\rm PN}=a_5' y^5\,,
\eeq
with
\beq
a_5'=-\frac{221}{6}+\frac{41}{32}\pi^2=-24.1879026944\,. 
\eeq
We then expect $a_2(u)$ to monotonically decrease towards $-\infty$ as $u$ varies between $0$ and $1/3$.
One can heuristically try to guess the way $a_2(u)$ will interpolate between the leading-order PN behavior \eqref{a24PN} and the LR behavior \eqref{a2LR}
by considering the doubly rescaled function 
\beq
b_2(y) \equiv \frac{(1-3y)^3}{y^5} a_2(y)\,.
\eeq  
As $y$ varies between $0$ and $\frac13$, the function 
$b_2(y)$ varies between $b_2(0)=a_5'\simeq -24.1879$ and $b_2(\frac13)=- \zeta^2 3^5/256 \simeq - 0.9597$. If we assume (as is the
case for the corresponding doubly-rescaled 1SF function $b_1(y) \equiv (1-3y)^{\frac12} a_1(y)/y^3$, see Ref. \cite{Akcay:2012ea}) that $b_2(y)$
varies (modulo its known $\sim y \ln y$ piece) roughly linearly in the interval $[0, \frac13]$, i.e. $b_2(y)\simeq a'_5 + y \, (c_2 - \frac{144}{5} \ln(3 y))$,
we can estimate its (logarithmically-corrected) slope $c_2$ as being $c_2 \simeq c_2^g$, with $c_2^g =  3 (b_2(\frac13)- a'_5) \simeq 69.7$. 
In other words, a guesstimate of the global strong-field behavior of $a_2(y)$ is
\beq
\label{a2g}
a_2^g(y)= \frac{y^5 \left(a_5' + y \, \left[ c_2^g -\frac{144}{5} \ln (3 y ) \right]\right)  }{(1-3y)^3}\,.
\eeq
The PN expansion of this guesstimate, namely
\beq
a_2^g(y)= a_5' y^5 +\left( 9 \, a_5' +c_2^g  -\frac{144}{5}\ln (3 y )\right) y^6 +\ldots \, ,
\eeq
suggests that the numerical value of the first unknown coefficient of $a_2(y)$, i.e., $a_6'$, might be of order 
${a'_6}^g=  9 \, a_5 +c_2^g    -\frac{144}{5}\ln 3 \simeq -179.6$. This result is not to be
taken as a firm numerical estimate, but only as an indication that the value of $a'_6$ can be reasonably expected to be of
order $- 200$.

\section{Conclusions}

Let us summarize our main results.

We have shown how EOB theory (together with the first law of binary dynamics) yields a simple, exact expression for the dependence of the redshift $z=z_1$ of a (nonspinning) mass $m_1$, in circular orbit around a nonspinning mass $m_2$, on the EOB gravitational potential $u=(m_1+m_2)/R$, in terms of the main radial EOB function $A(u;\nu)$, see Eq. \eqref{z1u}.
Using the latter expression, together with standard results of EOB theory, we derived in Eq. \eqref{z1_y_fin}  
the explicit relation between the second-order redshift function $z_{2\rm SF}(y)$ (defined by Eq. \eqref{zexp}) and the $O(\nu^2)$ contribution $a_2(u)$ to the EOB $A(u;\nu)$ potential (Eq. \eqref{Aexp}). Eq. \eqref{z1_y_fin} shows how to algebraically compute the function $a_2(\cdot )$ from $z_{2\rm SF}(y)$ (and a knowledge of $a_1(u)$).
For the convenience of the self-force community, we have also given the explicit relations between the various avatars [$z_{2\rm SF}(y)$, $\widehat z_{2\rm SF}(y)$, $U_{2\rm SF}(y)$, $\widehat U_{2\rm SF}(y)$] of the second-order redshift, see Eqs. \eqref{widehatz1andU_y_fin}, \eqref{avatars}.  

After recalling the remarkable cancellations taking place in the $\nu$-dependence of $A(u;\nu)$ (which starts being nonlinear in $\nu$ only at the 4PN level), we have considered the PN expansion of the second-order redshift and separated it into known and unknown parts. We emphasized that the known part (written in Eq. \eqref{z2K}) goes even beyond the 4PN level, as it includes the 5PN logarithm and the 5.5PN contribution. We expect that the known part $z_{2\rm SF}^{\rm known}(y)$, Eq. \eqref{z2K}, will give a good fit  of the data over a large range of frequency parameter $y$. We suggest to then interpret the upcoming 2SF data in terms of the difference $z_{2\rm SF}^{\rm numerical}(y)-z_{2\rm SF}^{\rm known}(y)$ (or some of its avatars) so as to directly extract the unknown parameters $a_6'$, $a_7'$ and $a_{7.5}'$ entering $z_{2\rm SF}^{a_2-\rm unknown}(y)$, Eq. \eqref{z2unK}.
Indeed, the parameters $a_6'$, $a_7'$ and $a_{7.5}'$ (and their higher-order analogs) are those of most direct dynamical relevance because they directly parametrize the PN expansion of the $O(\nu^2)$ contribution, $a_2(u)$, to the EOB radial $A$ potential. 

When going beyond the PN regime and exploring the strong field behavior of $z_{2\rm SF}(y)$ one will need, according to Eq. \eqref{z1_y_fin}, 
to use an accurate global analytic representation of the function $a_1(\cdot )$ in order to compute and subtract the $a_1$-dependent contributions to $z_{2\rm SF}(y)$.
We recall in this respect that such accurate global analytic representations were given in Section II B of Ref. \cite{Akcay:2012ea} (notably model 14 there).

We finally speculated on the LR singular behavior of both the various redshift functions and of $a_2(u)$. [We leave to future work the 2SF generalization
of the study of Ref. \cite{Akcay:2012ea}, namely the construction of a non-Damour-Jaranowski-Sch\"afer-gauge version of the EOB Hamiltonian that is
explicitly regular at $u=\frac13$.] 
We expect Eqs. \eqref{ULR} and \eqref{ZLR} to hold for the fractional redshift functions and Eqs. \eqref{a2k} 
and \eqref{a2LR} to hold for the 2SF contribution  $a_2(u)$ to the EOB $A$ potential. We also expect $a_2(u)$ to monotonically decrease from $0$ to $-\infty$ as $u$ increases from $0$ to $\frac13$, roughly as the guesstimate $a_2^g(y)$, Eq. \eqref{a2g}, and with a 5PN coefficient $a'_6 \sim -200$.

Let us finally mention that while the relations linking $z(y;q)$ to $A(u;\nu)$ we derived above should have a general validity, their application to the real conservative dynamics of binary systems depends on the precise definition that will be made in the second-order self-force computations. 
As explained, e.g., in \cite{Damour:1995kt}, and recently, in the Appendix of \cite{Damour:2016abl},
we personally favor the usual Fokker-like definition  of conservative dynamics based on the iterative use of a time-symmetric Green-function.
We therefore recommend that, when computing the redshift, 
both the 1SF metric perturbation $h_{\mu \nu}^{\rm 1SF}$, and the 2SF one $h_{\mu \nu}^{\rm 2SF}$, be computed 
by using the {\it time-symmetric} Green-function $G_{\rm sym}$ (in the background spacetime). [As indicated in the Appendix of \cite{Damour:2016abl}, this choice
might avoid infrared problems; though this issue clearly deserves a study of its own.] It is not clear to us that the prescriptions stated in \cite{Pound:2014koa} coincide with this iterated-$G_{\rm sym}$ one, nor it is clear that they will define, in general, a Hamiltonian evolution.

\section*{Aknowledgments}
T.D. wishes to thank Steve Detweiler for several informative email exchanges, over the past years, about SF theory.
D.B. thanks the Italian INFN (Naples) for partial support
and IHES for hospitality during the development of
this project. All the authors are grateful to ICRANet for
partial support.

\appendix

\section{Higher-order PN terms in the $O(\nu)$ correction to the EOB main radial potential}

We explicitly give here the coefficients of the beginning of the PN expansion of the $O(\nu)$ EOB radial potential $a_1(u)$. They were obtained 
through the 9.5 PN level (i.e. through $u^{10.5}$) in Ref. \cite{Bini:2015bla}.  Soon after, Ref. \cite{Kavanagh:2015lva} computed the PN expansion of the related quantity $U_{1\rm SF}(y)$  through the 22.5PN level, i.e. through $u^{23.5}$ . Below, we reproduce the analytical results of Ref. \cite{Bini:2015bla}, and complete them (analytically for the $u^{11}$ term, and numerically beyond that) by transcribing the results of  Ref. \cite{Kavanagh:2015lva}
in terms of  $a_1(u)$. Up to order $O(u^{11})$ we list the analytical values of the coefficients $a_n$ of $a_1(u) = \sum_{n\ge 3} a_n(\ln u) u^n$ 
(appropriately decomposed into powers of $\ln u$, according to $a_n(\ln u) = a_n^c + a_n^{\ln{}} \ln(u)+a_n^{\ln^2{}} \ln^2(u)+ \cdots$). 
Beyond that order, we give their numerical values. 
\begin{eqnarray}
a_1(u) &=& a_3 u^3 + a_4 u^4 + (a_5^c+a_5^{\ln{}} \ln(u))u^5 \nonumber\\
&&+(a_6^c+a_6^{\ln{}} \ln(u))u^6+a_{6.5}u^{13/2}\nonumber\\
&& +(a_7^c+a_7^{\ln{}} \ln(u))u^7+a_{7.5} u^{15/2} \nonumber\\
&&+(a_8^c+a_8^{\ln{}} \ln(u)+a_8^{\ln^2{}} \ln^2(u))u^8\nonumber\\
&& +a_{8.5}  u^{17/2 } \ldots \,,
\end{eqnarray}
where ($\gamma$ denoting Euler's constant)
\begin{widetext}
\begin{eqnarray}
a_3 &=& 2 \nonumber\\
a_4 &=& \frac{94}{3} - \frac{41}{32}\pi^2  \nonumber\\
a_5^c&=& -\frac{4237}{60} + \frac{128}{5}\gamma +\frac{2275}{512}\pi^2  + \frac{256}{5}\ln 2\nonumber\\
a_5^{\ln{}} &=& \frac{64}{5}\nonumber\\
a_6^c&=& -\frac{1066621}{1575} - \frac{14008}{105}\gamma + \frac{246367}{3072}\pi^2 
   - \frac{31736}{105}\ln 2 + \frac{243}{7}\ln 3  \nonumber\\
a_6^{\ln{}}&=& - \frac{7004}{105}\nonumber\\
a_{6.5}&=& \frac{13696}{525}\pi\nonumber
\end{eqnarray}
\begin{eqnarray}
a_7^c&=&   -\frac{1360201207}{907200} - \frac{5044}{405}\gamma   +\frac{608698367}{1769472}\pi^2 
 - \frac{2800873}{262144}\pi^4 + \frac{206740}{567}\ln 2 - \frac{4617}{14 }\ln 3 \nonumber\\
a_7^{\ln{}}&=& - \frac{2522}{405}  \nonumber\\
a_{7.5}&=& -\frac{512501}{3675}\pi\nonumber\\
a_8^c&=&     -\frac{187619320956191}{12224520000} + \frac{14667859963}{5457375}\gamma - \frac{109568}{525}  \gamma^2  +\frac{1836927775597}{2477260800}\pi^2   
+\frac{830502449}{16777216}\pi^4 \nonumber\\
&& +\frac{19361011651}{5457375}\ln 2
-\frac{438272}{525} \gamma \ln 2 -\frac{438272}{525}\ln^2 2 + \frac{3572343}{3520}\ln 3 
 +\frac{1953125}{19008} \ln 5  +\frac{2048}{5} \zeta(3) \nonumber\\
a_8^{\ln{}}&=&  
 \frac{14667859963}{10914750}  - \frac{109568}{525}\gamma  - \frac{219136}{525}\ln 2
   \nonumber\\
a_8^{\ln^2{}}&=&  - \frac{27392}{525}   \nonumber
\end{eqnarray}
\begin{eqnarray}
a_{8.5}&=& \frac{70898413}{6548850}\pi\nonumber\\
a_9^c &=& \frac{3121123440903397043}{8899450560000} -\frac{1198510638937}{198648450}\gamma + \frac{10894496}{11025}\gamma^2  
-\frac{53276112149251}{92484403200}\pi^2 
-\frac{23033337928985}{6442450944}\pi^4 \nonumber\\ 
&&-\frac{11647126988311}{993242250}\ln 2
+\frac{17379776}{3675}\gamma \ln 2
+\frac{322400}{63}\ln^2 2
+\frac{325284577623}{71344000}\ln 3 
-\frac{37908}{49}\gamma \ln 3 
 \nonumber\\
&& -\frac{37908}{49}\ln2 \ln3 
-\frac{18954}{49}\ln^2 3-\frac{2283203125}{1482624}\ln 5
 - \frac{152128}{105}\zeta(3) \nonumber\\
a_9^{\ln{}}&=& - \frac{1193425238617}{397296900} +\frac{10894496}{11025}\gamma   + \frac{8689888}{3675}\ln 2   - 
 \frac{18954}{49}\ln 3   \nonumber\\
a_9^{\ln^2{}} &=& \frac{2723624}{11025}  \nonumber\\
a_{9.5}^c &=& \frac{3008350528127363}{1048863816000} \pi - \frac{23447552}{55125}\gamma \pi  +\frac{219136}{1575}\pi^3 
 -\frac{46895104}{55125}\pi \ln 2 \nonumber\\
a_{9.5}^{\ln{}} &=& - \frac{11723776}{55125}\pi \nonumber
\end{eqnarray}
\begin{eqnarray}
a_{10}^c &=& \frac{75437014370623318623299}{18690753201120000}-\frac{21339873214728097}{1011404394000}\gamma  
+\frac{200706848}{280665}\gamma^2 +\frac{11980569677139}{2306867200}\pi^2 \nonumber\\
&&-\frac{24229836023352153}{549755813888}\pi^4
+\frac{27101981341}{100663296}\pi^6 +\frac{18605478842060273}{7079830758000}\ln 2 
-\frac{60648244288}{9823275}\gamma \ln 2  \nonumber\\
&&-\frac{121494974752}{9823275}\ln^2 2  
-\frac{6236861670873}{125565440}\ln 3 
+ \frac{360126}{49}\gamma \ln 3
 + \frac{360126}{49}\ln 2 \ln 3 \nonumber\\
&&
+\frac{180063}{49}\ln^2 3
+\frac{1115369140625}{124540416}\ln 5 
 +\frac{96889010407}{277992000 } \ln 7
 - \frac{1619008}{405}\zeta(3) \nonumber\\
a_{10}^{\ln{}} &=&  -\frac{21275143333512097}{2022808788000} + \frac{ 200706848}{280665}\gamma  
 -\frac{30324122144}{9823275}\ln 2 +\frac{180063}{49}\ln 3  \nonumber\\
a_{10}^{\ln^2{}} &=& \frac{50176712}{280665}    \nonumber 
\end{eqnarray}

\begin{eqnarray}
a_{10.5}^c&=& -\frac{185665618769828101}{24473489040000}\pi  
+\frac{2414166668}{1157625}\gamma \pi -\frac{5846788}{11025}\pi^3 
+\frac{377443508}{77175}\pi \ln 2 -\frac{246402}{343}\pi \ln 3 
\nonumber\\
a_{10.5}^{\ln{}}&& \frac{ 1207083334}{1157625}\pi \nonumber
\end{eqnarray}

\begin{eqnarray}
a_{11}^c&=& \frac{281895583614608101484671915254799}{9261923135147127244800000}
+\frac{730364677485317711340883}{6023874000444300000}\gamma
-\frac{1114681526261048}{49165491375}\gamma^2 \nonumber\\&&
+\frac{187580416}{165375}\gamma^3 
-\frac{9456705011234922635335117}{58656715985387520000}\pi^2 
-\frac{46895104}{33075} \gamma \pi^2 
-\frac{403529198843481822483991}{1662461581197312000}\pi^4\nonumber\\&&
-\frac{69677806640785}{12884901888}\pi^6
-\frac{220067102483775234280409}{6023874000444300000}\ln 2 
-\frac{2153292970969072}{49165491375}\gamma \ln 2 
+\frac{375160832}{55125}\gamma^2 \ln 2\nonumber\\&&
-\frac{93790208}{33075}\pi^2 \ln 2
+\frac{12035069804168}{49165491375}\ln^2 2
+\frac{750321664}{55125}\gamma \ln^2 2
+\frac{1500643328}{165375}\ln^3 2  \nonumber\\&&
+\frac{24590323035362369781}{167887271552000}\ln 3 
-\frac{99500270319}{4404400}\gamma \ln 3 
-\frac{99500270319}{4404400}\ln 2 \ln 3 
-\frac{99500270319}{8808800}\ln^2 3  \nonumber\\&&
+\frac{1361651238912109375}{139874081098752}\ln 5 
-\frac{7548828125}{2038608}\gamma \ln 5
-\frac{7548828125}{2038608}\ln 2 \ln 5 
-\frac{7548828125}{4077216}\ln^2 5 \nonumber\\&&
-\frac{5135117551571}{727056000}\ln 7
+\frac{228271533856 }{5457375}\zeta(3)
-\frac{3506176}{525}\gamma \zeta(3)
-\frac{7012352}{525}\ln2 \zeta(3)
-\frac{ 32768}{5}\zeta(5) 
\nonumber\\
a_{11}^{\ln{}}&=& \frac{733055111724601862700883}{12047748000888600000}  -
\frac{1114681526261048}{49165491375}\gamma+
\frac{93790208}{55125}\gamma^2 -\frac{23447552}{33075}\pi^2 \nonumber\\
&& -\frac{1076646485484536}{49165491375}\ln2
+\frac{375160832}{55125}\gamma \ln 2
+\frac{375160832}{55125}\ln^2 2
-\frac{99500270319}{8808800}\ln 3 \nonumber\\&&
-\frac{7548828125}{4077216}\ln 5   - \frac{1753088}{525}\zeta(3) 
  \nonumber\\
a_{11}^{\ln^2{}} &=& - \frac{278670381565262}{49165491375}+\frac{46895104}{55125}\gamma +\frac{93790208}{55125}\ln 2
\nonumber\\
a_{11}^{\ln^3{}} &=&\frac{23447552}{165375}  
\end{eqnarray}

\begin{eqnarray}
a_{12}^c&=&           -136026.4054204446524
\nonumber\\
a_{12}^{\ln{}}&=&     +92069.97011303800064
  \nonumber\\
a_{12}^{\ln^2{}} &=&  +2680.094911141314771 
\nonumber\\
a_{12}^{\ln^3{}} &=&  -575.6670078825180866
\end{eqnarray}

\begin{eqnarray}
a_{12.5}^c&=&     +411359.19012666159295       
\nonumber\\
a_{12.5}^{\ln{}}&=&    - 89983.79213560018813   
  \nonumber\\
a_{12.5}^{\ln^2{}} &=&  + 2723.4741165892301940   
\end{eqnarray}

\begin{eqnarray}
a_{13}^c&=&   -952605.909056261233         
\nonumber\\
a_{13}^{\ln{}}&=&  - 115325.0931536713629     
  \nonumber\\
a_{13}^{\ln^2{}} &=&  + 37380.38544992471130  
\nonumber\\
a_{13}^{\ln^3{}} &=&   - 958.1805076759092401
\end{eqnarray}

\begin{eqnarray}
a_{13.5}^c&=&   +610918.2394464063138         
\nonumber\\
a_{13.5}^{\ln{}}&=&  + 170007.28146400969249    
  \nonumber\\
a_{13.5}^{\ln^2{}} &=&  - 
 11672.693750604578136  
\end{eqnarray}

\begin{eqnarray}
a_{14}^c&=&     +1.350385599543487136\, \times10^6       
\nonumber\\
a_{14}^{\ln{}}&=&      - 984953.8855218083405 
  \nonumber\\
a_{14}^{\ln^2{}} &=&   - 
 19619.48449939532511 
\nonumber\\
a_{14}^{\ln^3{}} &=&   + 10567.846716039821126\nonumber\\
a_{14}^{\ln^4{}} &=&- 
 288.96957869200590289
\end{eqnarray}

\begin{eqnarray}
a_{14.5}^c&=&  -3.491894369332324840\, \times 10^6          
\nonumber\\
a_{14.5}^{\ln{}}&=&   + 843807.4189216130254   
  \nonumber\\
a_{14.5}^{\ln^2{}} &=&  - 
 14756.376627382853115   
\end{eqnarray}

\begin{eqnarray}
a_{15}^c&=&   + 8.453411734068058935\, \times 10^6        
\nonumber\\
a_{15}^{\ln{}}&=&   + 683868.4007401311506    
  \nonumber\\
a_{15}^{\ln^2{}} &=&  - 
 412898.58847750633693  
\nonumber\\
a_{15}^{\ln^3{}} &=&   + 2964.26039443923512\nonumber\\
a_{15}^{\ln^4{}} &=&+ 
 905.1064154160732068
\end{eqnarray}

\begin{eqnarray}
a_{15.5}^c&=&   -6.611739690898423300\, \times 10^6         
\nonumber\\
a_{15.5}^{\ln{}}&=&   - 2.5331466827392461057\, \times 10^6     
  \nonumber\\
a_{15.5}^{\ln^2{}} &=&  + 
 364363.73734313256846   
\nonumber\\
a_{15.5}^{\ln^3{}} &=&   - 7400.932837461527130
\end{eqnarray}

\begin{eqnarray}
a_{16}^c&=&      +5.49699806422373791\, \times 10^6      
\nonumber\\
a_{16}^{\ln{}}&=&      + 1.0474811240373705937\, \times 10^7 
  \nonumber\\
a_{16}^{\ln^2{}} &=&    - 
 284536.2707838998270
\nonumber\\
a_{16}^{\ln^3{}} &=&  - 92972.01464401640953
\nonumber\\
a_{16}^{\ln^4{}} &=&    + 
 2895.3944298738406392
\end{eqnarray}

\begin{eqnarray}
a_{16.5}^c&=&   2.0729749508779631714\, \times 10^7         
\nonumber\\
a_{16.5}^{\ln{}}&=&   - 8.154020295913721777\, \times 10^6    
  \nonumber\\
a_{16.5}^{\ln^2{}} &=&  - 
 320537.0424196010014  
\nonumber\\
a_{16.5}^{\ln^3{}} &=&    + 25457.964244814043318
\end{eqnarray}

\begin{eqnarray}
a_{17}^c&=&   -8.87075365872505874\, \times 10^7         
\nonumber\\
a_{17}^{\ln{}}&=&  + 6.53467956637541131\, \times 10^6      
  \nonumber\\
a_{17}^{\ln^2{}} &=&  + 
 4.546958839453220304\, \times 10^6   
\nonumber\\
a_{17}^{\ln^3{}} &=& - 
 126284.68507716340683 
\nonumber\\
a_{17}^{\ln^4{}} &=&   - 18108.698289996480898
\nonumber\\
a_{17}^{\ln^5{}} &=&     + 
 471.15801782925152928
\end{eqnarray}

\begin{eqnarray}
a_{17.5}^c&=&  1.3328203214222427101\, \times10^8        
\nonumber\\
a_{17.5}^{\ln{}}&=&  + 8.00320980641150146\, \times 10^6    
  \nonumber\\
a_{17.5}^{\ln^2{}} &=&    - 
 3.638624210501010330\, \times 10^6 
\nonumber\\
a_{17.5}^{\ln^3{}} &=&    + 62962.90624124632430
\end{eqnarray}

\begin{eqnarray}
a_{18}^c&=&    -2.515498325513126188\, \times 10^8        
\nonumber\\
a_{18}^{\ln{}}&=&   - 1.1922894900625442669\, \times 10^8   
  \nonumber\\
a_{18}^{\ln^2{}} &=&   + 
 1.0579211770929948012\, \times 10^7  
\nonumber\\
a_{18}^{\ln^3{}} &=& + 
 891056.9812725867700  
\nonumber\\
a_{18}^{\ln^4{}} &=&   - 33453.12599335842544
\nonumber\\
a_{18}^{\ln^5{}} &=&   - 
 877.4591663348057912   
\end{eqnarray}

\begin{eqnarray}
a_{18.5}^c&=&   -1.49229993389840279\, \times 10^7       
\nonumber\\
a_{18.5}^{\ln{}}&=&  + 1.0826017023366581713\, \times 10^8      
  \nonumber\\
a_{18.5}^{\ln^2{}} &=& + 
 3.111550184556815693\, \times 10^6  
\nonumber\\
a_{18.5}^{\ln^3{}} &=&     - 919803.3548556051975\nonumber\\
a_{18.5}^{\ln^4{}} &=& +15083.805973493017199
\end{eqnarray}

\begin{eqnarray}
a_{19}^c&=&  1.3538994200047574224\, \times 10^9         
\nonumber\\
a_{19}^{\ln{}}&=&    - 2.889719712681862193\, \times 10^8   
  \nonumber\\
a_{19}^{\ln^2{}} &=& - 
 4.805576848042135119\, \times 10^7   
\nonumber\\
a_{19}^{\ln^3{}} &=&  + 
 4.902006395972128981\, \times 10^6 
\nonumber\\
a_{19}^{\ln^4{}} &=&   + 70302.4657962266172
\nonumber\\
a_{19}^{\ln^5{}} &=&      - 
 5734.729153968339980
\end{eqnarray}

\begin{eqnarray}
a_{19.5}^c&=&    -2.3342132389583935641\, \times10^9      
\nonumber\\
a_{19.5}^{\ln{}}&=&  + 2.393706795331284037\, \times10^8     
  \nonumber\\
a_{19.5}^{\ln^2{}} &=& + 
 3.6103823234129382673\, \times10^7  
\nonumber\\
a_{19.5}^{\ln^3{}} &=&    - 493641.800817309027\nonumber\\
a_{19.5}^{\ln^4{}} &=& - 
 34430.30372113713688
\end{eqnarray}

\begin{eqnarray}
a_{20}^c&=&      +4.741091102573305324\, \times 10^9     
\nonumber\\
a_{20}^{\ln{}}&=&     + 1.3699923944647061573\, \times 10^9  
  \nonumber\\
a_{20}^{\ln^2{}} &=&  - 
 2.4804841985017927423\, \times 10^8  
\nonumber\\
a_{20}^{\ln^3{}} &=&  - 
 6.327070486932058680\, \times 10^6 
\nonumber\\
a_{20}^{\ln^4{}} &=&   + 767650.0821300041884
\nonumber\\
a_{20}^{\ln^5{}} &=&      + 
 16977.498788619257365\nonumber\\
a_{20}^{\ln^6{}} &=&  - 640.17660835212588741
\end{eqnarray}

\begin{eqnarray}
a_{20.5}^c&=&   -1.554500661703189789\, \times10^9     
\nonumber\\
a_{20.5}^{\ln{}}&=&    - 2.1444618040107173195 \, \times10^9  
  \nonumber\\
a_{20.5}^{\ln^2{}} &=&   + 
 1.6744038651919131540\, \times 10^8 
\nonumber\\
a_{20.5}^{\ln^3{}} &=& + 
 6.582343452113525271\, \times 10^6  \nonumber\\
a_{20.5}^{\ln^4{}} &=&   - 161073.36734648425473
\end{eqnarray}

\begin{eqnarray}
a_{21}^c&=&     -1.787411730448829770\, \times 10^{10}     
\nonumber\\
a_{21}^{\ln{}}&=&    + 5.800988660688537286\, \times 10^9 
  \nonumber\\
a_{21}^{\ln^2{}} &=&    + 
 4.437300025876944598\, \times 10^8 
\nonumber\\
a_{21}^{\ln^3{}} &=&  - 
 1.0018051535708298980\, \times 10^8  
\nonumber\\
a_{21}^{\ln^4{}} &=&    + 
 849802.7174561433836 
\nonumber\\
a_{21}^{\ln^5{}} &=&   + 95720.37725716444938  \nonumber\\
a_{21}^{\ln^6{}} &=&   + 
 80.8773788384730796
\end{eqnarray}

 \begin{eqnarray}
a_{21.5}^c&=&   +3.6533784749333873098\, \times 10^{10}      
\nonumber\\
a_{21.5}^{\ln{}}&=&   - 7.483988454959527808\, \times 10^9   
  \nonumber\\
a_{21.5}^{\ln^2{}} &=&   - 
 3.967252325467843392\, \times 10^8  
\nonumber\\
a_{21.5}^{\ln^3{}} &=&  + 
 2.0664713306648244037\, \times 10^7 \nonumber\\
a_{21.5}^{\ln^4{}} &=&  + 
 1.5339413832396392174\, \times 10^6 \nonumber\\
a_{21.5}^{\ln^5{}} &=&- 24593.786501542900423
\end{eqnarray}

\begin{eqnarray}
a_{22}^c&=&         -6.794231766690240157\, \times 10^{10} 
\nonumber\\
a_{22}^{\ln{}}&=&    - 1.4832318979201238365\, \times 10^{10}  
  \nonumber\\
a_{22}^{\ln^2{}} &=&   + 
 4.853892142268538180\, \times 10^9   
\nonumber\\
a_{22}^{\ln^3{}} &=&  - 
 1.872791394623178296\, \times 10^7 
\nonumber\\
a_{22}^{\ln^4{}} &=&    - 
 2.3298747461553891630\, \times 10^7 
\nonumber\\
a_{22}^{\ln^5{}} &=&  + 
 439152.9597858148096  \nonumber\\
a_{22}^{\ln^6{}} &=&   + 7050.010586529653912 
\end{eqnarray}

\begin{eqnarray}
a_{22.5}^c&=&   + 4.136798852425625556\, \times 10^{10}     
\nonumber\\
a_{22.5}^{\ln{}}&=&   + 3.417761112164835265\, \times 10^{10}  
  \nonumber\\
a_{22.5}^{\ln^2{}} &=&   - 
 5.217121376529257940\, \times 10^9 
\nonumber\\
a_{22.5}^{\ln^3{}} &=&  + 
 7.87671342117040392\, \times 10^6  \nonumber\\
a_{22.5}^{\ln^4{}} &=&  + 
 4.246280713165495896\, \times 10^6\nonumber\\
a_{22.5}^{\ln^5{}} &=& + 17232.59503122874741
\end{eqnarray}

\begin{eqnarray}
a_{23}^c&=&   + 2.0876199827245193609\, \times 10^{11}      
\nonumber\\
a_{23}^{\ln{}}&=&    - 9.877379696419447709\, \times 10^{10}  
  \nonumber\\
a_{23}^{\ln^2{}} &=&  - 
 2.409824624855166301\, \times 10^9    
\nonumber\\
a_{23}^{\ln^3{}} &=&   + 
 1.8428886435419457671\, \times 10^9
\nonumber\\
a_{23}^{\ln^4{}} &=&    - 
 4.962255033001139702\, \times 10^7  
\nonumber\\
a_{23}^{\ln^5{}} &=&    - 
 1.633843349737029919\, \times 10^6 \nonumber\\
a_{23}^{\ln^6{}} &=&    + 5490.221511483946715 \nonumber\\
a_{23}^{\ln^7{}} &=&   + 
 745.56622686995885663 
\end{eqnarray}

\begin{eqnarray}
a_{23.5}^c&=&   -4.925626220427599571\, \times 10^{11}      
\nonumber\\
a_{23.5}^{\ln{}}&=&   + 1.4219755003150905491\, \times 10^{11}   
  \nonumber\\
a_{23.5}^{\ln^2{}} &=&    + 
 8.109363034798849684\, \times 10^9 
\nonumber\\
a_{23.5}^{\ln^3{}} &=& - 
 1.4646378502431013409\, \times 10^9   \nonumber\\
a_{23.5}^{\ln^4{}} &=&  + 
 1.098747329777871858\, \times 10^7 \nonumber\\
a_{23.5}^{\ln^5{}} &=& + 245908.5910071982883
\end{eqnarray}

\end{widetext}

\end{document}